\documentclass[prl,twocolumn,showpacs,groupedaddress]{revtex4}
\usepackage{graphicx}
\usepackage{amsmath,amssymb}
\graphicspath{{./img/}}
\renewcommand{\section}[1]{{\par\it #1.---}}
\usepackage{hyperref}
\begin{document}
\title{Coherence stabilization of a two-qubit gate by AC fields}
\author{Karen M. \surname{Fonseca Romero}}
\altaffiliation[On leave from ]{Departamento de F\'\i sica,
Universidad Nacional, Bogot\'a, Colombia.}
\author{Sigmund Kohler}
\author{Peter H\"anggi}
\affiliation{Institut f\"ur Physik, Universit\"at Augsburg,
        Universit\"atsstra\ss e~1, D-86135 Augsburg, Germany}
\date{\today}
%
\begin{abstract}

We consider a CNOT gate operation under the influence of quantum bit-flip
noise and demonstrate that ac fields can change bit-flip noise into phase
noise and thereby improve coherence up to several orders of magnitude while
the gate operation time remains unchanged.  Within a high-frequency
approximation, both purity and fidelity of the gate operation are studied
analytically.  The numerical treatment with a Bloch-Redfield master
equation confirms the analytical results.


\end{abstract}
\maketitle


Despite the remarkable experimental realization of qubits
\cite{Nakamura1999a, Vion2002a, Chiorescu2003a} and two-qubit gates
\cite{Yamamoto2003a} in condensed matter systems, the
construction of a working quantum computer remains an elusive goal, not
only due to deficiencies of the control circuitry, but also due to the
unavoidable coupling to the environment.  Several proposals to overcome the
ensuing decoherence have been put forward, such as the use of
decoherence free subspaces \cite{Palma1996a, Duan1997a, Zanardi1997a,
Lidar1998a, Beige2000a}, quantum Zeno subspaces \cite{Facchi2002a},
dynamical decoupling \cite{Viola1999a, Vitali2002a, Gutmann2004a,
Falci2004a}, and coherent destruction of tunneling \cite{FonsecaRomero2004a}.

A single qubit under the influence of bit-flip noise can be modeled by the
spin-boson Hamiltonian
$
H_0 = -\frac{\Delta}{2}\sigma^z + \sigma^x \xi ,
$
where $\sigma^{x,z}$ denotes Pauli matrices and $\xi$ is a shorthand
notation for the quantum noise specified below.  The influence of the bath
is governed by the spectral density of the noise at the tunneling frequency
$\Delta/\hbar$.
A possible driving field may couple to any projection $\vec n$ of the
(pseudo) spin operator $\vec\sigma$, i.e., be proportional to $\vec n\cdot
\vec\sigma$. In Ref.~\cite{FonsecaRomero2004a}, two particular choices have
been studied and compared against each other:
A driving of the form $H(t)=A\sigma^z\cos(\Omega t)$ commutes with the
static qubit Hamiltonian while it modifies the bath coupling $\sigma^x\xi$
in such a way that the spectral density of the bath at multiples of the
driving frequency becomes relevant.  For a proper driving amplitude, this
eliminates noise with frequencies below the driving frequency which, thus,
should lie above the cutoff frequency of the bath.  This scheme represents
a continuous-wave version of dynamical decoupling \cite{FonsecaRomero2004a}.
By contrast, a driving of the type $H(t)=A\sigma^x\cos(\Omega t)$ renders
the qubit-bath coupling unchanged but renormalizes the tunnel splitting
$\Delta$ towards smaller values and thereby causes the so-called coherent
destruction of tunneling (CDT) \cite{Grossmann1991a, Grifoni1998a}.  Then,
decoherence is determined by the spectral density of the bath at a lower
effective tunnel frequency.  For an ohmic bath being linear in the
frequency, the consequence is that both decoherence and the coherent
oscillations in the rotating frame are slowed down by the same factor
\cite{FonsecaRomero2004a}.  Therefore, the number of coherent oscillations
is not enlarged and, thus, for single-qubit operations, CDT might be of
limited use.

In this work, we propose a coherence stabilization scheme for a CNOT gate
based on isotropic Heisenberg interaction \cite{Loss1998a, Kane1998a}.
Our scheme does not suffer from the drawbacks mentioned
above because (i) it involves only intermediately large driving frequencies
that can lie well below the bath cutoff and (ii) it does not increase the
operation time of the gate.  Since we shall employ a field that couples to
the same coordinate as the quantum noise, the present coherence
stabilization is different from dynamical decoupling.

\section{CNOT gate with bit-flip noise}
We consider a pair of qubits described by the Hamiltonian \cite{Loss1998a,
Kane1998a, Makhlin2001a, Thorwart2002a}
\begin{equation}
  \label{H0}
  H_\mathrm{gate}
 =\frac{1}{2} \sum_{j=1,2}
  \left( \Delta_j \sigma^x_j + \epsilon_j \sigma^z_j \right)
  + J\, \vec{\sigma}_{1}\cdot\vec{\sigma}_{2},
\end{equation}
with a spin-spin coupling of the Heisenberg type, where $j$ labels the
qubits.  In order to construct a quantum gate, the tunnel splittings
$\Delta_j$, the biases $\epsilon_j$, and the spin-spin coupling $J$ have to
be controllable in the sense that they can be turned off and that their
signs can be changed.
Then, a suitable sequence of interactions yields the CNOT operation
\cite{Loss1998a, Kane1998a, Makhlin2001a, Thorwart2002a}
\begin{equation}
  \label{eq:cnot1}
  U_\mathrm{CNOT}^H  \sim U_{H}(\pi/8)\, U_{1z}(\pi)\, U_{H}(\pi/8)\, ,
\end{equation}
where $U_{1z}(\varphi) = \exp(-i\varphi \sigma_1^{z}/2 )$ represents a
rotation of qubit 1 around the $z$-axis and $U_H(\varphi) = \exp(-i
\varphi\, \vec{\sigma}_1 \cdot\vec{\sigma}_2 )$ describes the time evolution
due to the qubit-qubit interaction.  The symbol $\sim$ denotes equality
up to local unitary transformations, i.e., transformations which act on
only one qubit.  Since we focus on decoherence during the
stage of the qubit-qubit interaction, we take as a working hypothesis that
one-qubit operations perform ideally.  Thus, decoherence takes place while
$\Delta_j = \epsilon_j = 0$ and $J>0$ during the required total qubit-qubit
interaction time $t_J = \pi\hbar/4J$.

The bit-flip noise is specified by the system-bath Hamiltonian
$H=H_\mathrm{gate}+H_\mathrm{coupl}+H_\mathrm{bath}$ where
$H_\mathrm{coupl} = \frac{1}{2} \sum_{j=1,2}\sigma_j^x \sum_\nu \hbar c_\nu
(a_{j\nu}^\dagger+a_{j\nu})$ describes the coupling of qubit $j$ to a
bath of harmonic oscillators with frequencies $\omega_\nu$ and
$H_\mathrm{bath} = \sum_{j\nu} \hbar\omega_\nu a_{j\nu}^\dagger a_{j\nu}$.
The bath is fully specified by its spectral density $I(\omega) =
\pi\sum_\nu c_\nu^2 \delta(\omega-\omega_\nu)$.  Within the present
work, we consider the so-called ohmic spectral density $I(\omega) =
2\pi\alpha\omega \exp(-\omega/\omega_c)$ with the dimensionless
coupling strength $\alpha$ and the cutoff frequency $\omega_c$.
In order to complete the model, we specify the initial condition for the
total density matrix to be of the Feynman-Vernon type, i.e., initially, the
bath is in thermal equilibrium and uncorrelated with the system,
$\rho_\mathrm{tot}(t_0) = \rho(t_0)\otimes
R_\mathrm{bath,eq}$, where $\rho$ is the reduced density operator of the
two qubits and $R_\mathrm{bath,eq}\propto\exp(-H_\mathrm{bath}/k_BT)$ is
the canonical ensemble of the bath.

If $\alpha k_BT$ is larger than the typical system energy $J$ and the
dissipation strength is sufficiently small, $\alpha\ll 1$, the dissipative
system dynamics is well described within a Born-Markov approach.  There,
one starts from the Liouville-von Neumann equation $i \hbar \dot
\rho_\mathrm{tot} = [H,\rho_\mathrm{tot}]$ for the total density operator
and obtains by standard techniques the master equation \cite{Kohler1997a}
\begin{align}
\dot\rho
= {} & \frac{1}{i\hbar}[H_\mathrm{gate},\rho]
  -\sum_j [\sigma_j^x, [Q_j(t),\rho ] ]
  - \sum_j [\sigma_j^x, \{ P_j(t),\rho \} ]
\nonumber \\
  \equiv {}& \frac{1}{i\hbar}[H_\mathrm{gate},\rho]
  - \Lambda(t)\rho
\label{eq:Born-Markov}
\end{align}
with the anti-commutator $\{A,B\}=AB+BA$ and
\begin{equation}
  \label{Q(t)}
  Q_j(t) = \frac{1}{4\pi}\int_0^\infty d\tau \int_0^\infty d\omega\,
  \mathcal{S}(\omega) \cos(\omega\tau) \widetilde\sigma_j^x(t-\tau,t) .
\end{equation}
Here, $\mathcal{S}(\omega) = I(\omega)\coth(\hbar\omega/2k_BT)$ is the
Fourier transformed of the symmetrically-ordered equilibrium bath
correlation function $\frac{1}{2}\langle\{ \xi(\tau),\xi(0)
\}\rangle_\mathrm{eq}$ of the collective coordinate $\xi=\sum_\nu c_\nu
(a_\nu^\dagger+a_\nu)$.  The notation $\widetilde X(t,t')$ is a shorthand
for the Heisenberg operator $U^\dagger(t,t') X U(t,t')$ with $U$ being the
propagator of the coherent system dynamics.
Note that $\mathcal{S}(\omega)$ and $I(\omega)$ are the same for both
qubits due to the assumption of two identical environments.
Replacing in Eq.~\eqref{Q(t)} the term $\mathcal{S}(\omega)\cos(\omega\tau)$
by $I(\omega)\sin(\omega\tau)$, yields the operator $P_j(t)$.  We emphasize
that the particular form \eqref{eq:Born-Markov} of the master equation is
valid also for an explicitly time-dependent Hamiltonian.

The heat baths, whose influence is described by the second and third term of
the master equation \eqref{eq:Born-Markov}, lead to decoherence, i.e., the
evolution from a pure state to an incoherent mixture.  Decoherence can be
measured by the decay of the purity $\operatorname{tr}(\rho^2)$ from the
ideal value 1.  The gate purity (later refered to as ``purity'')
$ \mathcal{P}(t) = \overline{\operatorname{tr} (\rho^2(t))} $,
which characterizes the gate independently of the specific input, results
from the ensemble average over all pure initial states \cite{Poyatos1997a}.
For weak dissipation, the purity is determined by its time derivative
at initial time. 
Performing the ensemble average, one finds the gate purity decay
\begin{equation}
  \label{eq:puritydecay}
  \dot{\mathcal{P}}(t)\big|_{t=0}
  = -2\,\overline{\operatorname{tr}(\rho\Lambda\rho)}
  = - \frac{2}{d(d+1)} \operatorname{Tr} \Lambda ,
\end{equation}
where $d=4$ is the dimension of the system Hilbert
space and $\mathop{\operatorname{Tr}}$ denotes the trace in
superoperator space defined as $\operatorname{Tr} \Lambda = \sum_{a,a'}
\operatorname{tr}( |a'\rangle\langle a| \Lambda |a\rangle\langle a'|)$
with $\{|a\rangle\}$ being an arbitrary orthonormal basis of the system
Hilbert space.
It can be shown that the term containing $P_j$ does not contribute
to $\operatorname{Tr}\Lambda$.

In order to evaluate for our setup the purity decay, we need explicit
expressions for the operators $Q_j$ and, thus, have to compute the
Heisenberg operators $\widetilde\sigma(t-\tau,t)$ for the Hamiltonian
$H_0=J\vec\sigma_1\cdot\vec\sigma_2$.  This calculation is most
conveniently done in the eigenbasis of $H_0$, i.e.\ in the basis of the
total (pseudo) spin $\vec L = \frac{1}{2}(\vec\sigma_1 + \vec\sigma_2)$.
After performing the time and the frequency integration in
Eq.~\eqref{Q(t)}, we finally arrive at
$
  \dot{\mathcal{P}}
  = -\frac{2}{5}\{ \mathcal{S}(0) + \mathcal{S}(4J/\hbar) \} ,
$ 
where we have ignored Lamb shifts and defined $\mathcal{S}(0) \equiv
\lim_{\omega\to 0} \mathcal{S}(\omega) = 4\pi\alpha k_BT/\hbar$. In
particular, we find that for low temperatures, $k_BT \lesssim J$,
decoherence is dominated by $\mathcal{S}(4J/\hbar)$ such that
$\dot{\mathcal{P}} \approx -16\pi\alpha J/5\hbar$.  This value reflects the
influence of the so-called quantum noise which cannot be reduced by further
cooling.

\section{AC driving field}
In order to manipulate the coherence properties, we act upon qubit~1 by an
ac field which causes a time-dependent level splitting according to
\begin{equation}
  \label{eq:driving}
  H_\mathrm{ac}(t) = f(t)\, \sigma_1^x ,
\end{equation}
where $f(t)$ is a $T$-periodic function with zero time-average.  Due to the
finite driving time, which acts only while the Heisenberg coupling
is switched on, the spectrum of the driving field acquires a dispersion
$\Delta\omega \approx \Omega/k$, where $k$ is the number of driving
periods.  To keep the influence of the dispersion small, we have to choose
$k\gg 1$.

Next, we derive within a high-frequency approximation analytical
expressions for both the coherent propagator $U(t,t')$ and the purity
decay~\eqref{eq:puritydecay}.  We start out by transforming the total
Hamiltonian into a rotating frame with respect to the driving by the unitary
transformation
\begin{equation}
\label{Uac}
U_\mathrm{ac}(t) = e^{-i\phi(t)\sigma_1^x},\quad
\phi(t) = \frac{1}{\hbar} \int_0^t dt'\,f(t') ,
\end{equation}
which yields the likewise $T$-periodic Hamiltonian $\widetilde H(t) =
U_\mathrm{ac}^\dagger(t) H_\mathrm{gate} U_\mathrm{ac}(t)$.  For large
driving frequencies $\Omega \gg J$, it is possible to separate time scales
and thereby replace $\widetilde H(t)$ by its time-average
\begin{equation}
\label{Heff}
\bar H = (J - J_\perp) \sigma_1^x\sigma_2^x
+ J_\perp \, \vec\sigma_1\cdot \vec\sigma_2 ,
\end{equation}
where the constant $J_\perp = J \langle \cos[2\phi(t)] \rangle_T$ denotes an
effective interaction ``transverse'' to the driving and
$\langle\ldots\rangle_T$ the time average over the driving period.
Within this approximation, the propagator of the \textit{driven} system reads
\begin{equation}
\label{Udriven}
U_\mathrm{eff}(t,t') = e^{-i\phi(t)\sigma_1^x}\, e^{-i\bar H(t-t')/\hbar}\,
e^{i\phi(t')\sigma_1^x} .
\end{equation}
Having this propagator at hand, we are in the position to derive explicit
expressions for the operators $\sigma_j^x(t-\tau,t)$ and $Q_j$ and,
consequently, also for the generator of the dissipative dynamics
$\Lambda$.  The calculation is conveniently done in the basis of the
total spin $\vec L$ and $L_x$ which, owing to the relation $\sigma_1^x
\sigma_2^x = \frac{1}{2}(\sigma_1^x+\sigma_2^x)^2-1$, is an eigenbasis of
the Hamiltonian \eqref{Heff}.  We insert the resulting expression for
$\Lambda$ into Eq.~\eqref{eq:puritydecay} and finally obtain the purity
decay
\begin{equation}
  \label{eq:purity2}
  \dot{\mathcal{P}}
  = -\frac{2}{5}\{ \mathcal{S}(0) + \mathcal{S}(4J_\perp/\hbar) \} .
\end{equation}
For $f(t)\equiv 0$, we find $J_\perp=J$ such that Eq.~\eqref{eq:purity2}
agrees with what we found in the static case; otherwise, the inequality
$|J_\perp| < J$ holds and, thus, the bath correlation function
$\mathcal{S}$ in Eq.~\eqref{eq:purity2} has to be evaluated at a lower
frequency.  For an ohmic or a super-ohmic bath, $\mathcal{S}(\omega)$ is a
monotonously increasing function and, consequently, the ac field reduces
$\dot{\mathcal{P}}$ (unless $J > \omega_\mathrm{cutoff}$).

The purity decay assumes its minimum for $J_\perp=0$.  This condition marks
the working points on which we shall focus in the following.  For an ohmic
spectral density $I(\omega) = 2\pi\alpha\omega$, the purity decay at the
working points becomes $\dot{\mathcal{P}} = -\frac{4}{5} \mathcal{S}(0) =
-8\pi\alpha k_BT/5$ for all temperatures.  This particular value has to be
compared to the purity decay in the absence of driving: An analysis reveals
that for $k_BT>J$, decoherence is essentially driving independent.  By
contrast for low temperatures, $k_BT<J$, the driving reduces the
decoherence rate by a factor $k_BT/2J$.  The reason for this
low-temperature behavior is that for $J_\perp=0$, the coupling operators
$\sigma_j^x$ commute with the effective Hamiltonian \eqref{Heff} such that
noise acts as pure phase noise with a strength proportional to the
temperature.

For a rectangular driving for which $f(t)$ switches between the values
$\pm A/2$, the condition $J_\perp=0$ yields $A=\hbar\Omega$ which corresponds
to the application of two $\pi$-pulses per period.
For a harmonic driving, $f(t)=A\cos(\Omega t)/2$, one obtains $J_\perp =
J J_0(A/\hbar\Omega)$, where $J_0$ denotes the zeroth-order Bessel function
of the first kind.  Then, at the working points $J_\perp=0$, the ratio
$A/\hbar\Omega$ assumes a zero of $J_0$, i.e., one of the values 2.405..,
5.520.., 8.654..,~\ldots.

So far, we focussed on decoherence and ignored the influence of the driving
on the coherent dynamics.  Since the driving changes also the coherent
dynamics, the pulse sequence of the CNOT operation needs a modification:
At the working points of the driven system, the propagator
\eqref{Udriven} becomes $U_\mathrm{eff}(t,t') = \exp[-i J \sigma_1^x
\sigma_2^x (t-t')/\hbar]$ and, thus, represents the time evolution caused
by a so-called Ising interaction $J \sigma_1^x \sigma_2^x$.  This time
evolution allows one to implement the alternative CNOT operation
$U_\mathrm{CNOT}^I \sim \exp( -i \pi \sigma_1^x \sigma_2^x /4 ) =
U_\mathrm{eff}(t+t_J,t)$ \cite{Gershenfeld1997a, Makhlin2001a}.  Note that the
interaction time $t_J = \pi\hbar/4 J$ is the same as for the original gate
operation $U_\mathrm{CNOT}^H$.  Since $U_\mathrm{ac}(2\pi/\Omega)$ is the
identity [cf.\ Eq.~\eqref{Uac}], we assume for convenience that the
driving period $2\pi/\Omega$ is an integer multiple of $t_J$, i.e.\ $\Omega
= 8kJ/\hbar$ with integer $k$ .

\section{Numerical solution}
The calculation of the purity decay presented above relies on a
high-frequency approximation which is correct only to lowest order in
$J/\hbar\Omega$.  Thus, these should be compared
to the exact numerical solution of the master equation
\eqref{eq:Born-Markov}.  An efficient scheme for that purpose is a modified
Bloch-Redfield formalism whose cornerstone is a decomposition into the
Floquet basis of the driven system \cite{Kohler1997a}:  According to the
Floquet theorem, the Schr\"odinger equation of a periodically driven
quantum system possesses a complete set of solutions of the form
$|\psi_\alpha(t)\rangle = \exp(-i\epsilon_\alpha t/\hbar)
|\phi_\alpha(t)\rangle$, where the so-called Floquet states
$|\phi_\alpha(t)\rangle$ obey the time-periodicity of the Hamiltonian and
$\epsilon_\alpha$ denotes the so-called quasienergy.
They are elements of an Hilbert space extended by a $T$-periodic time
coordinate and are computed from the eigenvalue equation
$[H(t)-id/dt]|\phi_\alpha(t)\rangle =
\epsilon_\alpha|\phi_\alpha(t)\rangle$.  In the Floquet basis
$\{|\phi_\alpha(t)\rangle\}$, the master equation
\eqref{eq:Born-Markov} obtains the form $\dot\rho_{\alpha\beta} =
-\frac{i}{\hbar}(\epsilon_\alpha-\epsilon_\beta) \rho_{\alpha\beta}
-\sum_{\alpha'\beta'} \Lambda_{\alpha\beta,\alpha'\beta'}\,
\rho_{\alpha'\beta'}$.  The benefit of this representation is that the time
dependence of the original Hamiltonian has been eliminated by choosing a
suitable basis.  Moreover, for weak dissipation, we can replace within a
rotating-wave approximation $\Lambda(t)$ by its time average
\cite{Kohler1997a}.  Finally, we integrate the master equation to obtain
the dissipative propagator which in turn allows to evaluate all quantities
of interest.

\begin{figure}[tb]
\includegraphics{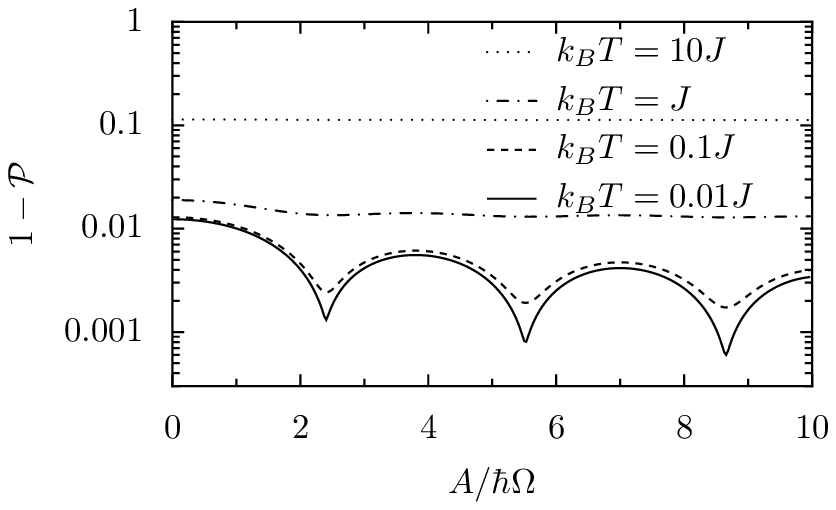}
\caption{Purity loss $1-\mathcal{P}$ during the Heisenberg interaction time
$t_J$ as a function of the driving amplitude for various temperatures.
The driving frequency is $\Omega=32J/\hbar$ and the dissipation strength
$2\pi\alpha=0.01$.  For $A=0$, the undriven situation is reproduced.
\label{fig:purity.A}}
\end{figure}
\begin{figure}[tb]
\includegraphics{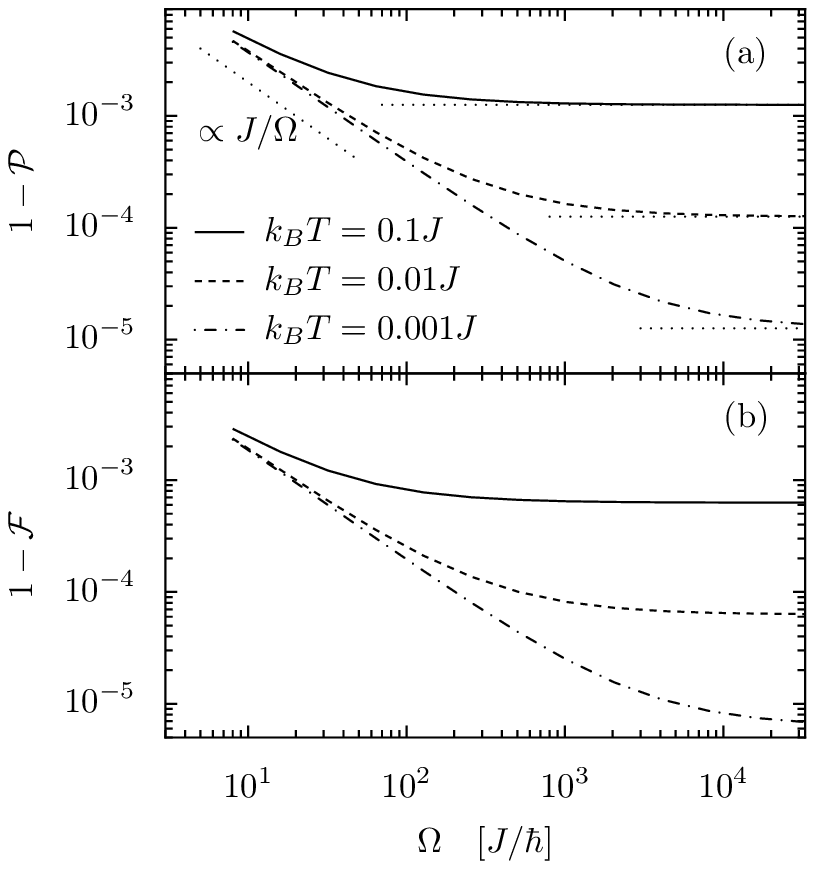}
\caption{(a) Purity loss during the Heisenberg interaction as a function of the
driving frequency.  The driving amplitude is adjusted such that
$1-\mathcal{P}$ assumes its first minimum; i.e.\ $A\approx 2.4\,\hbar\Omega$,
cf.\ Fig.~\ref{fig:purity.A}.
The dotted lines mark the analytical estimate $1-\mathcal{P}(t_J) \approx
-\dot{\mathcal{P}}(0)\,t_J$ and the proportionality to $J/\Omega$.
(b) Corresponding fidelity defect $1-\mathcal{F}$.
\label{fig:purityfidelity.omega}}
\end{figure}

%
In our numerical studies, we restrict ourselves to purely harmonic driving
$f(t)=A\cos(\Omega t)/2$.  The resulting purity loss during the interaction
time $t_J$ is depicted in Fig.~\ref{fig:purity.A}.  We find that for
$k_BT>J$, decoherence is fairly independent of the driving.  This behavior
changes as the temperature is lowered:  Once $k_BT<J$, the purity loss is
significantly reduced whenever the ratio $A/\hbar\Omega$ is close to a zero
of the Bessel function $J_0$.  Both observations confirm the preceding
analytical estimates.
The behavior at the first working point $A\approx 2.4\,\hbar\Omega$ is
depicted in Fig.~\ref{fig:purityfidelity.omega}(a).  For relatively low
driving frequencies, we find the purity loss being proportional to
$J/\Omega$.  There are significant deviations from the analytical result
for small $\Omega$ because the low-frequency regime is not within the
scope of our analytical treatment.  With increasing driving frequency,
the discrepancy decreases until finally thermal noise dominates
decoherence.  The numerical solution confirms almost perfectly the results
from the high-frequency approximation.

Still, there remains one caveat:
The gate operation $U_\mathrm{CNOT}^I$ relies on the fact that
$U_\mathrm{eff}$ is a good approximation for the dynamics of the driven
gate because any deviation represents a coherent error.  Therefore, we
still have to demonstrate that such coherent errors are sufficiently small.
As a measure for this, we employ the so-called fidelity
$
\mathcal{F}
= \overline{\operatorname{tr}[\rho_\mathrm{ideal}\,\rho(t_J)]}
$ \cite{Poyatos1997a}
which constitutes the overlap between the real outcome of the operation,
$\rho(t_J)$, and the desired final state $\rho_\mathrm{ideal} = U_I(\pi/4)
\rho_\mathrm{in} U_I^\dagger(\pi/4)$ in the average over all pure initial
states.  An ideal operation is characterized by $\mathcal{F}=1$.  Here,
$U_I(\varphi) = \exp(-i\varphi \sigma_1^x \sigma_2^x)$ denotes the
propagator due to an ideal Ising spin-spin interaction.
Figure \ref{fig:purityfidelity.omega}(b) shows the fidelity defect
$1-\mathcal{F}$ of the real gate operation at the first working point.  In
the relevant temperature regime $k_BT<J$, we find the crucial result that
the fidelity defect is of the same order or even smaller than the purity
loss.  Thus, we can conclude that coherent errors are not of a hindrance.

%
In summary, we have put forward a strategy for reducing decoherence in a
two-qubit quantum gate during the stage of qubit-qubit interaction.  This
coherence stabilization scheme is different from previous proposals in
two respects:
First, it does not rely on the manipulation of the coordinate that couples
the qubits to the environment.  By contrast, the central idea of our scheme
is rather to suppress the \textit{coherent} system dynamics ``transverse''
to this sensitive system coordinate.  In that way, the bit-flip noise,
which produces unwanted transitions, is turned into pure phase noise which
is proportional to the temperature.  This enables a coherence gain by
cooling.
The second difference is that the proposed scheme eliminates also the noise
coming from the spectral range above the driving frequency and, thus, is
particularly suited for ohmic noise spectra with large cutoff frequencies.
Moreover, the driven system still allows one to perform the desired CNOT
operation with high fidelity and within the same operation time as in the
absence of the control field.  Hence, the gained coherence time fully
contributes to the number of feasible gate operations.

%
This work has been supported by the Freistaat Bayern via the quantum
information initiative ``Quanteninformation l\"angs der A8'' and by the
Deutsche Forschungsgemeinschaft through SFB 631.


\end{document}